\documentclass[10pt,sigconf,letterpaper]{acmart}
\renewcommand\footnotetextcopyrightpermission[1]{} 
\begin{document}\sloppy

\title{System to Identify and Elide Superfluous JavaScript Code for Faster Webpage Loads}

\author{Utkarsh Goel}
\affiliation{\institution{Akamai Technologies, Inc.}}
\email{ugoel@akamai.com}

\author{Moritz Steiner}
\affiliation{\institution{Akamai Technologies, Inc.}}
\email{mosteine@akamai.com}

\maketitle

\pagenumbering{gobble}

\section*{Abstract}
Many websites import large JavaScript~(JS) libraries to customize and enhance user experiences.
Our data shows that many JS libraries are only partially utilized during a page load, and therefore, contain superfluous code that is never executed.
Many top-ranked websites contain up to hundreds of kilobytes of compressed superfluous code and a JS resource on a median page contains 31\% superfluous code.
Superfluous JS code inflates the page weight, and thereby, the time to download, parse, and compile a JS resource.
It is therefore important to monitor the usage and optimize the payload of JS resources to improve Web performance.
However, given that the webpage design and functionality could depend on a user's preferences or device, among many other factors, actively loading webpages in controlled environments cannot cover all possible conditions in which webpage content and functionality changes.
\looseness -1

\vspace{3pt}
\indent
In this paper, we show that passive measurement techniques, such as real user monitoring systems~(RUM), that monitor the performance of real user page loads under different conditions can be leveraged to identify superfluous code.
Using a custom man-in-the-middle proxy~(similar to a content delivery network's proxy server), we designed a systematic approach for website developers that relies on pages loaded by real users to passively identify superfluous code on JS resources.
We then elide any superfluous code from JS resources before subsequent page load requests.
Our data shows that eliding superfluous JS code improves the median page load time by 5\% and by at least 10\% for pages in the long tail.
Through results presented in this paper, we motivate for the need for rigorous monitoring of the usage of JS resources under different real world conditions, with the goal to improve Web performance.
\looseness -1



\section{Introduction}
\label{sec:introduction}
  
To generate aesthetically appealing Web experiences and to monitor page's performance, many websites bundle JavaScript~(JS) libraries as a development and deployment convenience~\mbox{\cite{bestjs}}.
JS bundles offer a large set of cross-browser functionality, for example via JS Polyfills, which when imported may remain under-utilized across difference webpages~\cite{polyfills}.
As such, a JS bundle that contains large amounts of superfluous code can slow down the page load process because not only does the superfluous JS code increase the page weight, it also increases the time to download, parse, and compile the JS resource before it can be executed~\cite{costofjs}.
We perform a measurement study to investigate the presence of superfluous code in 100+ top-ranked websites.
Our data shows that websites contain up to hundreds of kilobytes of compressed JS payload that never executes during a page load. 
Additionally, we observe that the median JS resource contains 31\% superfluous code.
\looseness -1

 \begin{figure}[t]
 \centering
 \minipage{0.49\textwidth}\includegraphics[width=\linewidth]{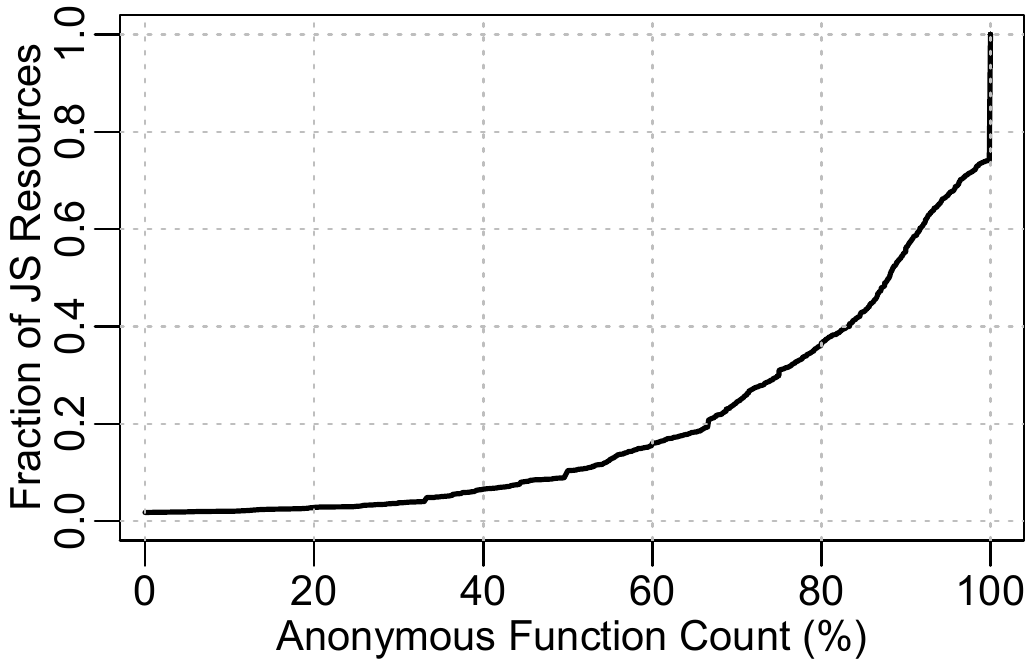}
 \vspace{-15pt}
\caption{CDF of anonymous function count.}
 \label{fig:anonymous_functions}
 \endminipage
 \vspace{-10pt}
\end{figure}

Given this large amounts of superfluous code on websites, we argue that website developers should perform rigorous monitoring of their JS resources to understand usage based on real user page loads, with the goal of eliding superfluous code before delivering to clients. 
Active experiments could be used to perform static analysis of JS resources via Chrome's Coverage API~\cite{coverageapi}, however, experiments performed using the Coverage API will have several limitations.
First, Web developers will need to load pages in controlled environments and therefore cannot trigger all website behaviors based on user's personal preferences, device's capabilities, screen size, phone model, geography, network quality, among many other factors~\cite{geo,rwd,im}.
As a result, the Coverage API cannot be used to capture JS usage under all real world conditions under which a website behavior may change~\cite{meenan}.
Second, the Coverage API does not capture the usage of anonymous function declarations in JS resources and reports all anonymous functions as superfluous.
As shown in Figure~\ref{fig:anonymous_functions}, the median JS resource contains 88\% anonymous function declarations.
In fact, all function declarations in top 20\% JS resources are anonymous.
As a result JS usage measurements performed using the Coverage API will be inaccurate.
Therefore, the Coverage API under-estimates the JS usage.
And third, the data captured by the Coverage API is not exposed to JS and therefore, RUM-based systems cannot collect data about the usage of JS resources.
\looseness -1

In this paper, we show that passive measurement techniques are better suited to estimate the usage of JS resources during page loads, similarly to how Real User Monitoring~(RUM) systems capture various Web performance metrics~\mbox{\cite{catchpoint,dynatrace}}.
A passive monitoring solution for JS resources can also be employed by website owners or content delivery networks~\cite{mpulse}. 
We make the following contributions in this paper.
\looseness -1

\vspace{5pt}
\noindent
\textbf{Proxy Server:} We designed an HTTP(S) proxy, similarly to CDN servers, that not only terminates TCP/TLS connections and serves HTTP(S) requests, but can also authoritatively modify JS resources on behalf of the website owner~\cite{akamai}.
For every requested JS resource, the proxy rewrites the resource payload modifying each function declaration so that the added snippet executes when the function containing it executes during the page load.
After page load, the modified JS resource sends back information containing which functions executed during the page load, which is then used to elide superfluous functions. 
To maintain page functionality in case an elided function needs to execute, the proxy adds logic to synchronously download and execute the elided function body without causing page errors.
\looseness -1

\vspace{0pt}
\noindent
\textbf{Measurements:} To the best of our knowledge, there is currently no known best-practice on how to reliably measure JS usage via passive measurement techniques. 
Our paper is the first academic measurement effort that investigates JS usage on 100+ top-ranked e-commerce, banks, auto manufacturers, media, and entertainment websites, randomly chosen from Alexa top 1000 websites.
We investigate JS usage on various first and third party resources, where we classify a resource as third party when it is not served by the first party infrastructure~\cite{GoelThird}.
We investigate differences in JS usage across two types of mobile devices: a slow, small sized screen mobile device~(Moto~G) and a relatively faster, bigger sized screen mobile device~(Moto~G4).
Finally, we perform experiments to learn the superfluous components of JS resources loaded on different webpages and measure the Web performance post superfluous JS elision.
Note that since mobile devices generally have slower CPUs than laptops and desktop machines, and since JS operations during page load happen on device's CPU, the impact of superfluous JS on Web performance is higher on slower devices~\cite{MoritzSmartphone15,wprof,costofjs}.
Therefore, we tailor our experiments to investigate the JS usage and performance improvements from elision on pages loaded on mobile devices.
\looseness -1

\vspace{3pt}
\noindent
\textbf{Inferences Drawn:} Using our custom proxy server and mobile devices on the WebPageTest platform for loading webpages~\cite{wpt}, we make the following observations.
First, many top-ranked websites contain up to hundreds of kilobytes of compressed superfluous JS code, which accounts to up to 38\% of total first party JS and up to 71\% of total third party JS on some pages.
Second, the median JS resource on a given webpage is 31\% superfluous.
Moreover, 48\% of the median JS library in the JQuery framework was found to be superfluous for the pages in our dataset -- suggesting that developers import the entire third party library but only ever use it partially.
Third, for 13\% of JS libraries loaded during our experiments, the set of executed JS functions varied based on the device's screen size and the User-Agent header passed in the HTTP requests.
And fourth, by eliding first-party superfluous JS, the median page load time improves by 5\% and at least 20\% for pages in the long tail.
\looseness -1

The rest of the paper is organized as follows.
Section~\ref{sec:experimental} describes our experimental methodology to identify and elide superfluous JS code.
In Sections~\ref{sec:resultsjs}, we discuss our results.
In Section~\ref{sec:related_work}, we discuss related work.
In Section~\ref{sec:discussion}, we discuss challenges with identifying and eliding superfluous CSS rules.
Finally, we conclude and provide future directions in Section~\ref{sec:conclusions}.
\looseness -1

\section{Experimental Methodology}
\label{sec:experimental}

The identification and elision of superfluous JS code works in two phases.
The first phase is the learning phase, where we collect data as to whether or not a function inside a JavaScript file executes during the page load. 
In particular, when the proxy serves a JS resource to the client, it makes several changes in the background for future requests of the same resource.
First, the proxy modifies the JS resource to declare an array at the top of the JS resource.
It then parses the resource to build an abstract syntax tree~(AST), using which it learns the beginning and the end byte ranges of every function declaration in the JS resource~\cite{ast}.
The proxy then iterates over function declarations, generates a unique identifier for each declaration, and adds a JS snippet~(as the first statement in the declaration).
This snippet inserts the function identifier to the array declared at the top of JS resource.
Note that this snippet executes whenever the function containing it executes during the page load.
\looseness -1

Finally, the proxy adds another JS snippet at the top of the JS resource that sends to our Web server the contents of the globally declared array after the page has loaded.
Website developers often wait for different page events, such as \texttt{window.onLoad}, to fire before they execute additional JS code for downloading functionality and content for the page~\cite{onload}.
Therefore, in our experiments the above snippet executes when the browser triggers the \texttt{window.loadEventEnd} event -- indicating that all listeners, such as \texttt{setTimeout}s or \texttt{Promise}s, attached to the \texttt{window.onLoad} event have finished executing~\cite{loadeventend}.
At \texttt{window.loadEventEnd}, the array containing the identifiers of all executed functions are sent to our server.
When our server receives the beacon containing identifiers of all executed functions, the proxy transitions into the second phase to elide superfluous functions.
\looseness -1

In the second phase, every function whose identifier was not present in the beacon is elided.
However, it is possible that certain JS functions execute under specific conditions not captured during the phase one. 
As a result, if such functions are deleted and are needed to execute during the page load, the webpage functionality will break. 
Therefore, instead of deleting the entire function declaration, the proxy replaces the body of each superfluous function with a stub that when executed, issues a synchronous \texttt{XMLHttpRequest}~(XHR) request to download the elided function body, followed by running the JS \texttt{eval} method on the downloaded function body and providing the method the appropriate context and function arguments~\cite{eval,xhr}.
When eliding superfluous functions, the proxy copies the function body into a separate file and its HTTP path on the server is hard-coded in the XHR request.
Note that the stub makes XHR requests in synchronous mode to strictly preserve the function execution order, in case multiple elided functions are called.
However, to ensure that we never elide a potentially needed function, the learning phase must be conservatively employed to identify JS functions that execute based under different scenarios.
Note that since we tailor our solution for website developers and CDN vendors and since both of them can only authoritatively modify JS resources that they serve, in the second phase, we elide superfluous JS from first party resources only.
\looseness -1

\looseness -1

 \begin{figure}[t]
 \centering
 \minipage{0.49\textwidth}\includegraphics[width=\linewidth]{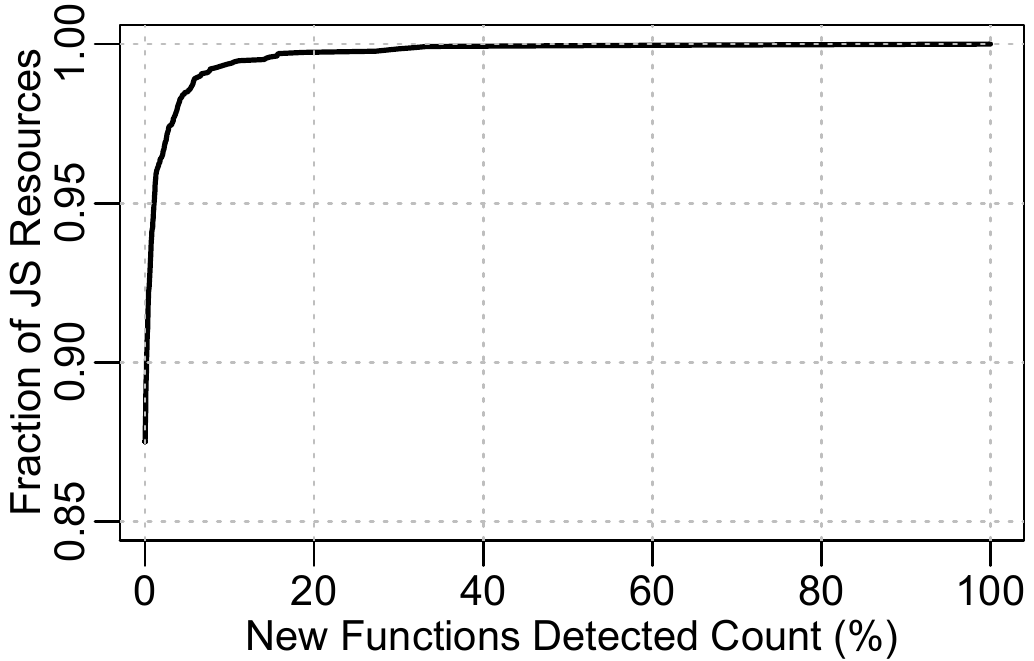}
 \vspace{-15pt}
\caption{CDF of newly detected function count.}
 \label{fig:requested_functions}
 \endminipage
 \vspace{-10pt}
\end{figure}

\vspace{5pt}
\noindent
\textbf{Client Setup:} To measure the performance differences in page loads with and without superfluous JS code, we utilize the WebPageTest~(WPT) platform and its client devices to load 100+ webpages~\cite{wpt}.
WPT has a fleet of 20 Moto~G and 20 Moto~G4 mobile devices located in Dulles, VA, USA.
Moto~G is small screen sized mobile device with a 1.2 GHz quad-core CPU.
Moto~G4 is bigger screen sized mobile device with a 1.5GHz octa-core CPU.
We utilized both of these devices to load pages using the Chrome browser~(v69) installed on these devices and to understand the JS usage that occur from small and large screen sized devices, as well as understand the impact of eliding superfluous JS on Web performance for page loaded over devices with different CPU clock speed. 
We deploy our proxy on an AWS EC2 instance in the same geographic region as the chosen WPT devices, to obtain similar effect as how clients connect to nearby CDN servers~\cite{akamai}.
\looseness -1

Next, we want the client devices to load webpage resources from our proxy, instead of website owners' servers, so as to modify JS resources.
Therefore, when running experiments, we provide Chromium browser the \texttt{--host-resolver-rules} flag that bypasses the DNS lookup process and uses the provided IP address to connect and download resources~\cite{hostresolverrules}.
In our experiments, this flag looks like \mbox{\texttt{MAP * <Proxy IP>}}, where \texttt{*} indicates all hostnames. 
Since in our experiments, our proxy is not authoritative for the requested webpage resources, for resources requested over TLS connections, the proxy returns a self-signed certificate.
We configure the Chromium browser to ignore any certificate errors by passing the \texttt{--ignore-certificate-errors} flag~\cite{ignorecertificateerror}.
Finally, to prevent resource loading from browser cache and affecting measurement data, we pass the \texttt{--incognito} flag to Chromium so that the pages are loaded in the incognito mode.
\looseness -1

Finally, we begin loading webpages in the learning phase to identify superfluous JS functions.
We use the WPT API to configure mobile devices and Chromium browser to load 100+ webpages, one at a time, for five times each.
In the learning phase, we load each page five times to detect any function execution that is triggered on a sampling basis, for example, due to A/B testing of new features.
Figure~\ref{fig:requested_functions} shows the distribution of the percentage of functions triggered randomly across five page loads. 
As shown in Figure~\ref{fig:requested_functions}, we observe that new function executions were detected in repetitive page loads for about 12\% of JS resources across all webpages.
\looseness -1

Note that since we load pages in Chrome's incognito mode, each time the browser finishes loading the page and quits the browser process, all cached data, including DNS entries, TCP/TLS connections, HTTP payload, is deleted. 
Also note that loading the same page multiple times increases the likelihood of executing functions that may run on a sampling basis -- thus triggering different code paths.
When the proxy receives HTTP requests, it forwards the request to the website owner's servers and stores a copy of the resource locally.
For any JS resource, the proxy modifies each function declaration~(as described earlier), stores a local copy of the modified JS resource, and sends it back to the client to trigger the learning phase for that JS resource.
Upon execution of this JS resource, the proxy receives a beacon containing function identifiers of all executed functions.
The proxy stores identifiers of all executed functions on a per JS resources basis in a local database.
When new function identifiers are found in a beacon from the five page loads, the proxy appends all new function identifiers to the database.
\looseness -1

Once the learning phase completes for a webpage, the proxy retrieves the list of executed function identifiers for each JS resource and elides all functions whose identifiers did not appear in the list. 
The proxy then stores a copy of the elided JS resource locally.
At the end of learning phase, the proxy has a copy of all original resources requested during the page load, a copy of the JS resources with added JS snippets, and a copy of JS resources with elided superfluous functions. 
Storing a local copy allows proxy to serve HTTP requests without forwarding requests to the website owner's server -- thus preventing results to be not impacted by any fluctuations in the network performance between the proxy and the website owner's server.
After a JS resource is elided, the proxy stores several characteristics about the JS resource in the database, such as the number of functions~(total and elided), number of anonymous functions~(total and elided), JS bytes~(total and elided).
\looseness -1

For each webpage for which we elide superfluous JS code, before starting performance measurements, we compared the webpage's visual and functional correctness with that of the original webpage.
Specifically, we compared for the error logs printed in the browser console, the number of requests, and manually checked for visual similarity. 
In all cases, we found the two versions of the webpage same in visual appearance and functionality.
\looseness -1

\section{Results}
\label{sec:resultsjs}

Now that the proxy has learned which part of the JS code is superfluous and has a copy of original and elided JS resources available to serve, we now load webpages to measure the performance differences with and without superfluous JS elision.
We iterate over our 100+ webpage list to load each page 20 times in the \texttt{original} mode and 20 times in the \texttt{elided} mode.
The \texttt{original} mode refers to a setup where the proxy sends original resources as retrieved from website owner's servers.
The \texttt{elided} mode refers to a setup where the proxy sends original resources, except for JS resources where the proxy sends a copy with elided superfluous code.
\looseness -1

 \begin{figure}[t]
 \centering
 \minipage{0.49\textwidth}
\includegraphics[width=\linewidth]{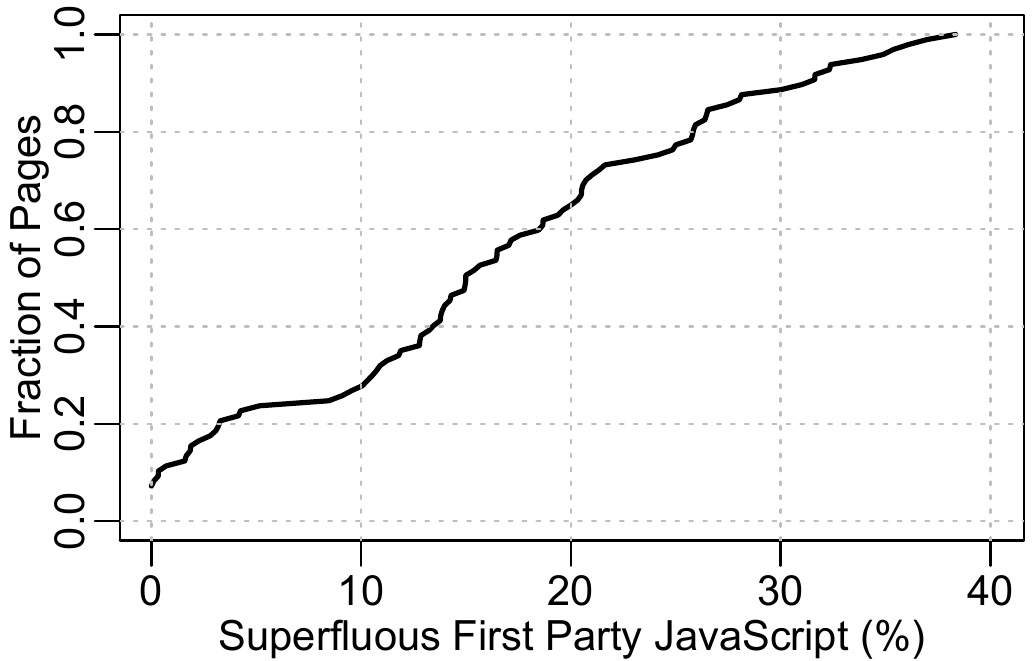}
 \vspace{-15pt}
\caption{CDF of superfluous JS (\%) on webpages.\looseness -1}
 \label{fig:fp_elidejsperc_cdf}
 \endminipage
 \vspace{-10pt}
\end{figure}

In Figure~\ref{fig:fp_elidejsperc_cdf}, we show the amount of first party superfluous JS code on 100+ webpages.
From the figure we observe that 16\% of first party JS code is superfluous on the median page and up to 38\% on other pages. 
Note that this superfluous JS code accounts to up to 700\,KiB of compressed JS payload, which could decompress to over a megabyte of JS code.
Before the browser can execute even a non-superfluous portion of the JS resource, all superfluous JS payload must also be downloaded, decompressed, and parsed -- impacting the page's performance.
We perform similar investigation for third party JS resources and found that for the median page 16\% of third party JS is superfluous; however, given many website developers import many JS frameworks to introduce additional functionality on the page, third party JS resources could be up to 71\% superfluous on some pages.
While website developers are not authoritative to make any changes to third party JS resources, however, if such resources could be arranged to be served from their first party infrastructure, superfluous code in third party JS resources could also be identified and thus elided.
\looseness -1

\begin{figure}[t]
 \centering
 \minipage{0.49\textwidth}\includegraphics[width=\linewidth]{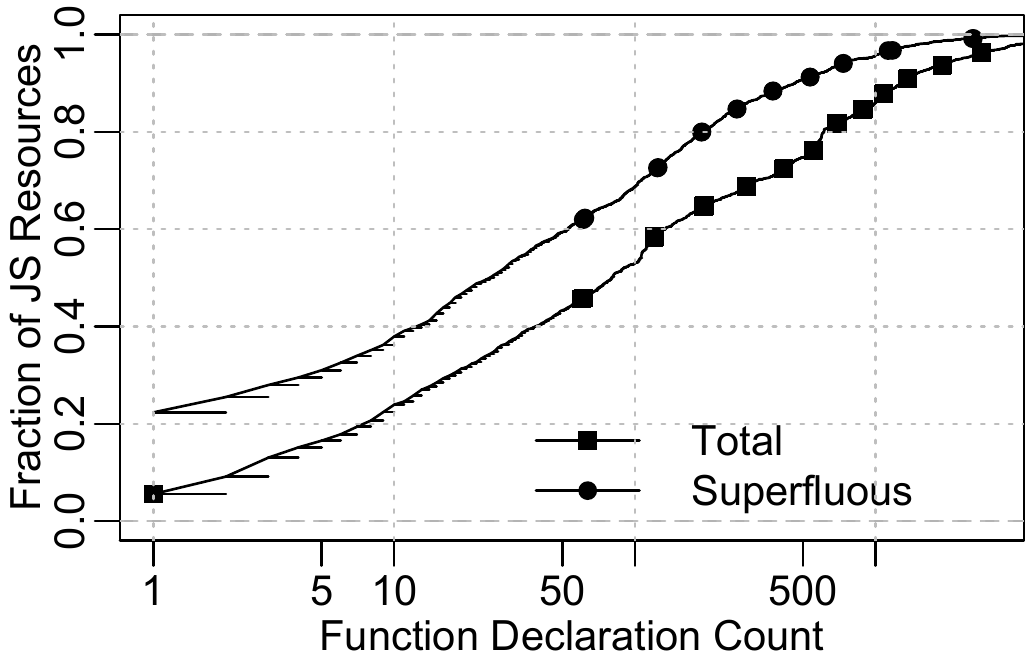}
 \vspace{-15pt}
\caption{CDF of total and superfluous functions.}
 \label{fig:total_vs_superfluous_functions}
 \endminipage
 \vspace{-10pt}
\end{figure}

Next, we investigate the number of superfluous functions on a per-resource basis, in over 2500 JS resources loaded across 100+ webpages.
Figure~\ref{fig:total_vs_superfluous_functions} shows that the median JS resource has a total of 80 function declarations, out of which 25~(31\%) of declarations are superfluous.
The JS resource at the 80th percentile contains at least 512 function declarations, out of which at least 140~(27\%) declarations are superfluous.
Exploring the collected data further~(but not shown on the graph), we found that the median library in the JQuery framework is 48\% superfluous.
Similarly, we found that in other frameworks, such as Adobe Tag Manager and Clicktale, the median library is 16\% and 29\% superfluous, respectively.
\looseness -1

\begin{figure}[t]
 \centering
 \minipage{0.49\textwidth}
\includegraphics[width=\linewidth]{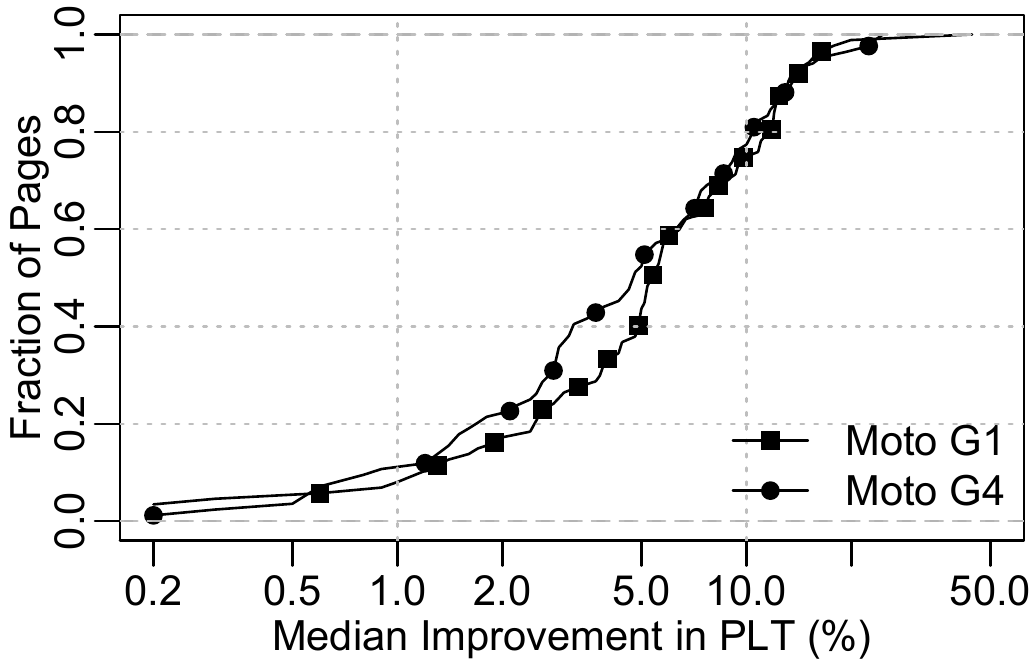}
 \vspace{-15pt}
\caption{CDF of the median percentage improvement in PLT. Notice the log scale on x-axis.}
 \label{fig:js_improvement_cdf}
 \endminipage
 \vspace{-10pt}
\end{figure}

Based on observations, we explore the opportunity of optimizing JS resources by eliding superfluous code.
Since mobile devices have slower CPUs than laptops and desktops, and since JS execution happens on the device's CPU, we measure the webpage performance with and without superfluous JS only on mobile devices.
In Figure~\ref{fig:js_improvement_cdf}, we show distributions of median percentage improvements in the page load time~(PLT) metric, as observed when loading pages on a fleet of Moto~G and Moto~G4 mobile devices.
Note that we refer PLT as the time since the start of the navigation until the browser triggers the \texttt{window.onLoad} event~\cite{nt}.
From the figure we observe that elision of superfluous JS improves the median PLT by 5\% for the median page and at least 10\% for the pages in the long tail.
Note that one should not compare the two distributions on the figure and speculate that performance improvements on slower phones are higher than that on a relatively faster phone.
While this assumption may be true if the pages were identical across page loads on the two devices.
However, we found that 13\% of JS libraries loaded during our experiments execute different set of JS functions on Moto~G and Moto~G4 mobile devices. 
Therefore, the pages loaded on the two devices have different amounts of superfluous code that was elided by our proxy.
\looseness -1

\section{Limitations and Future Work}
\label{sec:conclusions}

While our experiments are designed to passively monitor the JS usage on webpages, they do not capture execution of JS functions that triggers only from user-interactions with the page.
Therefore, our experiments may have over-estimated the amount of superfluous JS for certain pages.
We argue that a comprehensive coverage of all executed JS functions can only be achieved from a real world deployment of our methodology.
A real world deployment will not only help accurately identify all executed JS functions but also help identify how websites change JS behavior under different conditions, such as the geographic locations, time of day, etc.
Future work with real-world data could help address the current limitations of this paper.
\looseness -1

\section{Related Work}
\label{sec:related_work}

The Web performance community has developed several techniques to improve JS-related operations on client devices and improve page load times.
For example, Google Chrome implements the script-streaming thread that parses JS resources in parallel to their download -- reducing the parsing time and the PLT by up to 10\%~\cite{scriptstreaming}.
Current implementation of script-streaming thread parses exactly one JS resource at a time on the script-streaming thread, even when many JS resource download in parallel.
As such, the benefits of script-streaming thread remain limited and therefore, other JS resources on the page must be parsed after they are downloaded.
Our previous work shows that PLTs can be further improved by 6\% by rearranging script tags in the HTML in such a way that allows for larger JS resources to be parsed on the script streaming thread~\cite{goelscriptstreaming}.
\looseness -1

WebAssembly is a new programming language that is smaller than JS code and allows for faster execution than JS code~\cite{wa}.
The WebAssembly code is delivered to clients in an already parsed and compiled byte code format, unlike JS resources that must be parsed and compiled by the browser after they are downloaded~\cite{waperf1,waperf2}.
Similarly, BinaryAST is an under-development proposal by Mozilla and Facebook to speed up parsing and compilation operations of JS resources on Web browsers.
With BinaryAST, developers write JS code but convert it into a binary representation for sending it to a compatible Web browser~\mbox{\cite{ast,binaryast1,binaryast}}.
\looseness -1

Netravali~\textit{et\,al.} developed two tools, Scout and Polaris, to generate fine-grained dependency graphs of webpage resources with the goal of loading resources in the order they are needed on webpages.
Their experiments suggest an improvement of 34\% in the median PLT~\cite{polaris}. 
Another tool, Klotski, investigates improving user-perceived webpage performance by performing lexical analysis of the HTML document to learn the dependency tree~\cite{klotski}.
Netravali~\textit{et\,al.} also proposed a tool for Web servers to precompute JS heap and DOM trees pertaining to a given webpage with the goal to eliminate intermediary JS computations on the end-user device's CPU~\cite{prophecy}.
\looseness -1

To reduce the workload on devices with slow CPUs, such as mobile phones, several Web browsers improve webpage performance by performing CPU-intensive DOM, CSSOM, and the render tree manipulations in well-provisioned cloud data centers~\cite{operamini,puffin,shandian}.
Amazon's Silk browser allows webpages to be loaded through their cloud data centers, where some of the webpage operations are performed in the cloud and the results are delivered as compressed blob to the client device~\cite{amazonsilk,amazonsilkstudy1,amazonsilkstudy2}.
\looseness -1


Removal of superfluous code has also been studied from the perspective of improving security.
Azad~\textit{et\,al.} investigate the benefits of debloating various server-side Web applications~\cite{236200} and show both performance and security improvements by eliding the dead code on webpages.
Similarly to our experiments, their experiments do not detect the coverage of any code that executes due to user interactions. 
However, the methodology we discuss in this paper is specifically designed to work with existing RUM solutions and CDN proxy servers, which offers the benefit to learn JS coverage from various user-interactions through a real world deployment.
Moreover, the work by Azad~\textit{et\,al.} does not implement any fallback mechanisms to load function elided mistakenly, which as we discuss in the paper is critical for proper functionality of the webpage.
\looseness -1

 \begin{figure}[t]
 \centering
 \minipage{0.49\textwidth}
\includegraphics[width=\linewidth]{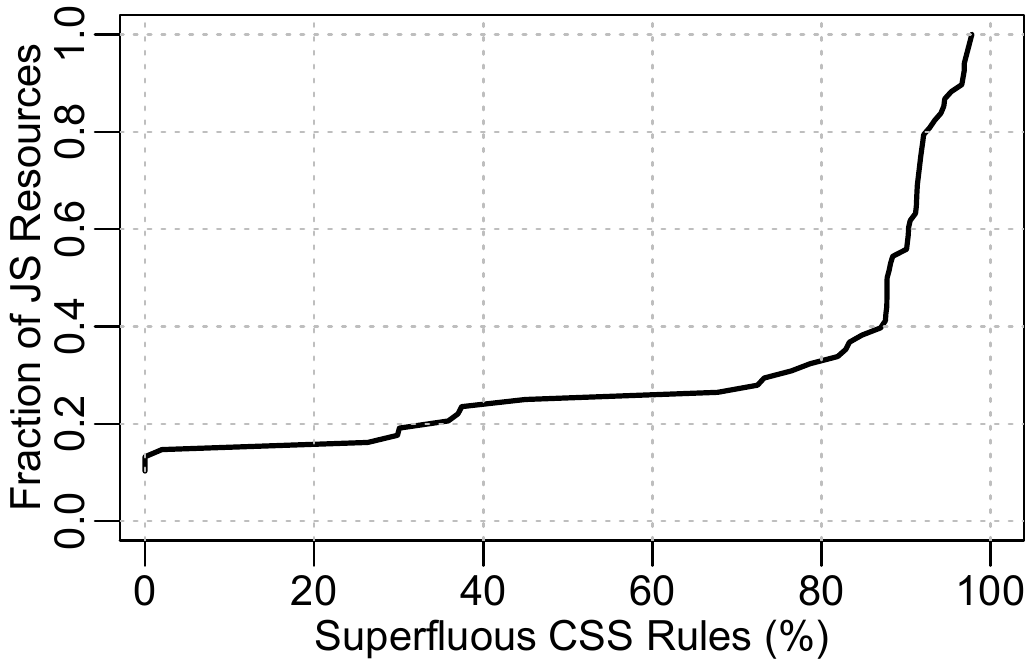}
 \vspace{-15pt}
\caption{CDF of superfluous CSS rules.}
 \label{fig:superfluous_css_perc}
 \endminipage
 \vspace{-10pt}
\end{figure}

\section{Challenges with CSS Elision}
\label{sec:discussion}

We also investigate the existence of superfluous Cascading Style Sheet~(CSS) rules on webpages.
To the best of our knowledge, Chrome Coverage API is the only known method to programmatically detect CSS usage on webpages.
Our analysis suggests 
the measurements done by the Coverage API are accurate, however, similarly to JS usage measurements, the accuracy depends on \emph{when} we perform the measurement.
As mentioned earlier, the Coverage API currently cannot be used to perform passive measurements and therefore, active experiments performed using it can only help collect CSS usage based on how webpages load apply CSS rules for conditions simulated in an in-lab setting.
\looseness -1

Unlike JS, CSS rules are key-value pairs and do not contain JS-like functions that can be modified to detect execution.
Moreover, Web browsers do not throw errors when a certain CSS rule is not found in the CSS Object Model~(CSSOM)~\cite{cssom}.
As such, there is no know mechanism to passively detect superfluous CSS rules, elide it, and have it back-filled in case its needed during the page load.
Therefore, if a non-superfluous CSS rule is elided, the resultant CSSOM could generate an incorrect page rendering.
As a result, it is challenging to systematically verify whether a page with no superfluous CSS rules will always produce correct UI for the end-user.
Knowing the challenges, we explore the opportunity of measuring the CSS usage and PLT improvements with elision, via active experiments using the Coverage API.
\looseness -1


When a webpage is loaded, the Coverage API provides byte ranges of CSS rules that were applied to the page.
We use the puppeteer library to actively load pages via our proxy and to fetch these byte ranges from the Coverage API.
We then elide the superfluous rules from the CSS resource that do not lie in that range.
Figure~\ref{fig:superfluous_css_perc} shows a surprising amount of superfluous CSS rules on many popular pages. 
Specifically, the CSS rules on the median page 87\% superfluous -- resulting in download of up to two megabytes of superfluous CSS payload on some pages.
\looseness -1

Note that since CSS rule parsing is not CPU-intensive~\cite{costofjs}, network conditions are more likely to impact the PLT.
Therefore, in these experiments we compare the PLT across different simulated network conditions, such as 3G and LTE.
We use the following network settings to simulate a 3G network between a WPT client and our proxy server: round trip time~(RTT) of 150\,ms, downlink bandwidth of 1.6\,Mbps, and uplink bandwidth of 768\,Kbps~\cite{wpttrafficshaper1,wpttrafficshaper2,wpttrafficshaper}.
Similarly, we simulate LTE network as follows: RTT of 70\,ms and  downlink and uplink bandwidth of 12\,Mbps each.
\looseness -1

\begin{figure}[t]
 \centering
 \minipage{0.49\textwidth}
\includegraphics[width=\linewidth]{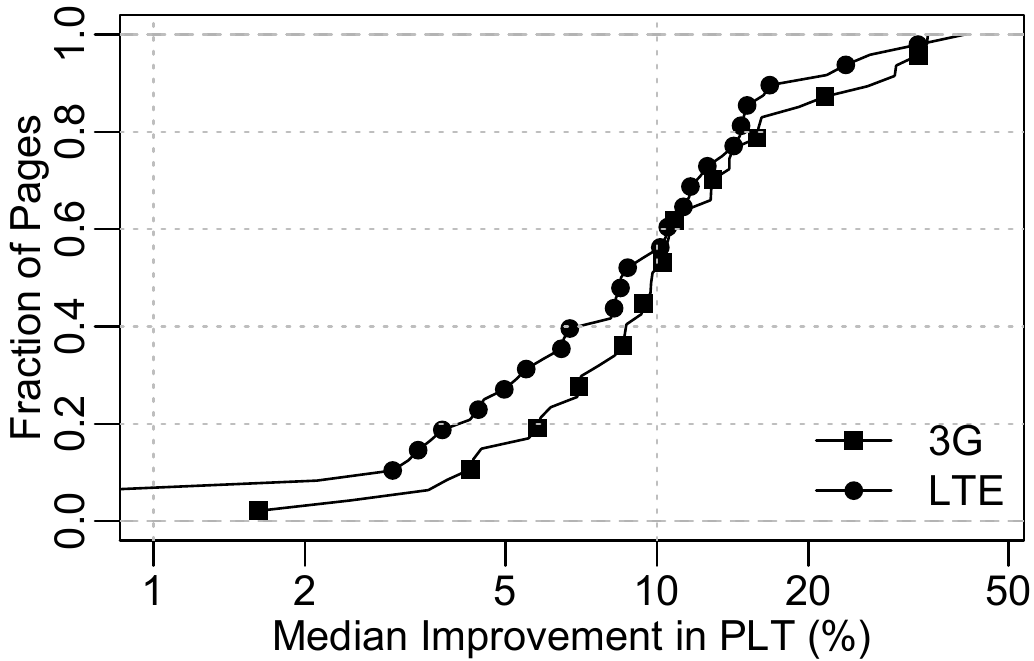}
 \vspace{-15pt}
\caption{CDF of the median percentage improvement in PLT. Notice the log scale on x-axis.}
 \label{fig:css_improvement_cdf}
 \endminipage
 \vspace{-10pt}
\end{figure}

In Figure~\ref{fig:css_improvement_cdf}, we show the PLT differences with and without superfluous CSS rules.
We observe that pages loaded under simulated 3G network conditions experience a slightly higher improvement in the median PLT than pages loaded under simulated LTE network for some webpages. 
Moreover, for the median page the PLT improves by 10\%. 
We also observe improvements up to 34\% under both 3G and LTE conditions.
Since at the moment there is no way to passively learn CSS usage on real user page loads, we are working with Chromium engineers to expose the CSS coverage data via JS method. 
\looseness -1

Finally, to allow further research and to maintain paper anonymity while its under the review process, we plan to open-source the proxy after acceptance of the paper.
\looseness -1

\section{Conclusions}
\label{sec:conclusions}

Website developers import JS library bundles to generate aesthetically appealing experiences for their users.
Our research shows that these libraries are often only partially utilized -- resulting in large amounts of superfluous code on the website that must be downloaded, decompressed, and parsed before execution.
When these JS operations block the main thread to the device's CPU to process the superfluous code, the PLT time gets negatively impacted.
In this paper, we present our early analysis on how superfluous JS code could be identified and elided from webpages using passive measurement techniques. 
Our results indicate median PLT improvements of 5\% and 10\% improvements for the $80^{th}$ percentile of pages.
\looseness -1

Our goal with the paper is to not only provide guidance on how developers could modify their JS resources to monitor usage but to motivate for the need of comprehensive monitoring of JS resources.
Based on our work, we recommend website developers to invest efforts in JS usage monitoring techniques, either on their web servers or via their contracted CDN vendors, with the goal to improve the end-user's Web experience.
\looseness -1

\section*{ACKNOWLEDGMENTS}
Thanks to Stephen Ludin and Martin Flack for providing feedback on an earlier version of this manuscript.
\looseness -1

\section*{Disclosure}
In the interest of full disclosure, Akamai has patents both granted and pending in this subject matter area.
As always, an understanding of the patent landscape is advisable before proceeding with any particular solution.
The positions, strategies, or opinions reflected in this article are those of the authors and do not necessarily represent the positions, strategies, or opinions of Akamai.

\bibliographystyle{abbrv}
\bibliography{elide}
\end{document}